\documentclass[9pt]{article}
\usepackage{amsfonts}
\usepackage{pstricks}
\usepackage{pst-node}
\usepackage{epsfig}
\usepackage{epsfig}
\usepackage[latin5]{inputenc}
\usepackage[T1]{fontenc}
\usepackage{lipsum}
\usepackage{amsmath}
\usepackage{authblk}
\usepackage{float}

\begin{document}
                \def\ba{\begin{eqnarray}}
                \def\ea{\end{eqnarray}}
                \def\w{\wedge}
                \def\d{\mbox{d}}
                \def\D{\mbox{D}}

\begin{titlepage}
\title{Colliding Gravitino Plane Waves in Simple Supergravity}
\author{Tekin Dereli${}^{1,2,}$\footnote{tdereli@ku.edu.tr, tekindereli@maltepe.edu.tr} , Yorgo \c{S}eniko\u{g}lu${}^{1,}$\footnote{yorgosenikoglu@maltepe.edu.tr}}
\date{%
    ${}^{1}$ \small Department of Basic Sciences, Faculty of Engineering and Natural Sciences, \\Maltepe University, 34857 Maltepe,\.{I}stanbul, Turkey\\%
    ${}^{2}$ \small Department of Physics, Ko\c{c} University, 34450 Sar{\i}yer, Istanbul,Turkey\\[2ex]%
    \today
}
\maketitle



\begin{abstract}

\noindent We give a family of exact solutions of $N=1$ supergravity field equations in $D=4$ dimensions that describe the collision of plane-fronted gravitino waves.


 \end{abstract}

\vskip 1cm



\end{titlepage}

\newpage

\section{Introduction}

\noindent Arguably one of the most far-reaching predictions of Einstein's theory of general relativity is the existence of gravitational waves as fluctuations of space-time curvature
that propagate with the speed of light  in free space. Indeed  it was  Einstein himself at first  who discussed  the expected physical properties of gravitational waves
in the weak field limit in a linearized approximation to the Einstein field equations. The relatively recent observational  evidence on the reality of gravitational waves
still makes essential use of the linearized approximation. On the other hand Einstein's gravitational field equations are intrinsically non-linear. Exact source-free Einstein
 solutions that describe individual gravitational waves are also known for a long time. In this context, the basic non-linearity of Einstein's field equations is most clearly manifested
 when two  plane-fronted gravitational waves collide face-to-face. There are two seminal papers by Khan and Penrose\cite{khan-penrose} and by Szekeres\cite{szekeres1}   that describe the collision of  impulsive gravitational waves in a distributional sense\cite{szekeres2} where an exact Einstein solution is given with an essential curvature singularity
 that develops during the collision due to the mutual focusing of two gravitational plane-fronted waves. This basic collision problem was generalized soon after by writing down exact solutions that describe the collision of gravitational and electromagnetic shock  waves in Einstein-Maxwell theory\cite{bell-szekeres} on the one hand, and by the inclusion of polarization effects on collision of gravitational waves\cite{nutku-halil,chandrasekhar-ferrari} on the other.  Related generalizations of the collision problems for massless gravitons, spin-1 and spin-0 bosons, dealing with all various possible combinations  of couplings between them, soon followed\cite{matzner-tipler}. Such solutions are complemented through discussions of the mathematical aspects of  collision problems in relativistic theories of gravitation in general\cite{yurtsever,clarke-hayward}. Nevertheless two cases concerning the coupling of massless spin-1/2 and/or spin-3/2 fermions to gravitational waves  remained
inconclusive and not solved so far.

The collision of neutrino plane waves in general relativity was first considered by Grifitths\cite{griffiths4,griffiths3,griffiths2}. He started out for the construction of an exact solution of the coupled Einstein-Weyl equations in a torsion-free space-time geometry using the NP formalism. But the collision of the so-called ghost neutrino waves cannot be extended explicitly to the interaction region and the problem remained open since then.
The interpretation of neutrino waves as a ghost configuration in a semi-Riemannian approach was later criticized by Griffths\cite{griffiths1} himself and
an extension of the Einstein-Weyl equations to space-times with torsion was given\cite{dereli-tucker} in the language of complex quaternion valued exterior differential forms
on space-time.  This extension though didn't lead to a consistent exact colliding neutrino wave solution either. But it
points towards  a close structural similarity between the field equations satisfied by a massless  spin-1/2 Weyl neutrino and a massless spin-3/2 Rarita-Schwinger gravitino where
the space-time torsion is specified in both cases algebraically in terms of quadratic expressions in the spinor fields.

The coupling  of  higher-spin massless Rarita-Schwinger spinors to Einstein's general relativity poses consistency problems in general, such as the non-causal gravitino propagation. It is remarkable that the $N=1$ simple supergravity field equations
in $d=4$ dimensions are free of these inconsistencies because of  local supersymmetry covariance\cite{deser-zumino}.
A family of exact plane wave solutions of simple supergravity was first given by Aichelburg and Dereli\cite{dereli-aichelburg}.
Another family of exact solutions was found later by Urrutia\cite{urrutia} for which
the space-time torsion and the gravitino stress-energy-momentum tensors vanish identically. Thus the simple supergravity field equations reduce
to vacuum Einstein field equations plus the massless  Rarita-Schwinger equation in that vacuum background.
Nonetheless, both AD and U families of supergravity plane waves consist of non-trivial solutions in the sense that they cannot be super-gauge generated from a vacuum Einstein configuration\cite{aichelburg-urbantke}.  In fact it was shown\cite{beler-dereli1} that gauge-generated supergravity plane wave solutions  are always found in a particular superposition
of AD-solutions and U-solutions.Therefore either family on its own will be non-trivial.
Recently we provided a re-formulation of exact supergravity plane wave solutions in terms
of complex  quaternion valued exterior differential forms\cite{dereli-senikoglu}.

The sole discussion of colliding plane waves in classical simple supergravity that is found in the literature\cite{rosenbaum} was given in the framework of U-solutions for which
the curvature in the interaction region is free of essential singularities. In what follows we construct a family of exact solutions that describe colliding gravitino
plane waves in the framework of AD-solutions. In Section:2 simple supergravity field equations are given in terms of complex quaternion valued exterior differential
forms over 4-dimensional space-time. The Majorana 4-spinors in this language correspond to minimal left ideals in the complex quaternion algebra
(See \cite{dereli-senikoglu} and the references therein).
The Bell-Szekeres metric together with a gravitino ansatz in null-coordinates in AD-form is presented in Section:3.
The real 4-component Majorana spinor equivalent of our gravitino ansatz is explicitly given in the Appendix.The corresponding  torsion and curvature 2-forms are evaluated and the simple supergravity field equations are reduced. An exact solution that describes the collision of two plane-fronted waves in supergravity that collide face-to-face is explicitly written out.
Both the metric and gravitino fields are consistently patched across the null-cone by implementing O'Brien-Synge boundary conditions\cite{obrien-synge}.
Section:4 contains some concluding comments.

\newpage

\section{Simple Supergravity Field Equations}

\noindent The basic supergravity multiplet involves a massless spin-2 graviton field associated with a metric tensor
\ba
g = \eta_{ab} e^a \otimes e^b
\ea
given in terms of an orthonormal co-frame $e^a = e^{a}_{\mu}(x)  dx^{\mu}$ in a local coordinate chart $\{x^{\mu}\}$.
The frame indices $a,b,\dots=0,1,2,3$ are raised and lowered through the Minkowski metric $\eta_{ab}=diag(-,+,+,+)$.
The curvature 2-forms are derived from the connection 1-forms
\ba
\omega^{a}_{\;\;b} = \hat{\omega}^{a}_{\;\;b} + K^{a}_{\;\;b}
\ea
where the Maurer-Cartan structure equations
 $de^a + \hat{\omega}^{a}_{\;\;b} \wedge e^b =0$   and  $K^{a}_{\;\;b} \wedge e^b = T^a $
determine uniquely the Levi-Civita connection 1-forms  $\{ \hat{\omega}^{a}_{\;\;b} \}$ of the metric and  the contortion 1-forms  $\{ K^{a}_{\;\;b}\}$ of the torsion, respectively. The superpartner of the graviton field is a massless spin-3/2 gravitino field that is  described by a Majorana 4-spinor valued 1-form \ba
\psi = \psi_a e^a = \psi_{\mu} dx^{\mu},
\ea
whose covariant exterior derivative is given by
 \ba
 \nabla \psi = d\psi + \frac{1}{2} \omega^{ab} \sigma_{ab} \wedge \psi
 \ea
 with the matrix generators of the local Lorentz group
$ \sigma_{ab} \equiv \frac{1}{4} \left [ \gamma_a , \gamma_b  \right ]$. Our conventions and the choice of $\gamma$-matrices are given below in the Appendix.

\medskip

\noindent The simple supergravity field equations whose solutions we will be seeking consist of the Einstein equations\footnote{$*$ to the left of a $p$-form denotes its
Hodge dual $(4-p)$-form where we set the volume 4-form $*1=e^0 \wedge e^1\wedge e^2 \wedge e^3$.}
\ba
-\frac{1}{2} R^{bc} \wedge *e_{abc} = \frac{i}{4} \left ( \bar{\psi} \wedge \gamma_5\gamma_a \nabla \psi \right ),
\ea
coupled to  the Rarita-Schwinger equation
\ba
i e^a \gamma_a \wedge \nabla \psi =0,
\ea
with both being subject to an algebraic constraint
\ba
T^a = \frac{i}{4} \left ( \bar{\psi} \wedge \gamma^a \psi \right )
\ea
 that fixes the space-time torsion 2-forms.

\medskip

\noindent  We now give a brief description of the 4-dimensional space-time geometry
in the language of complex quaternion valued exterior differential forms.  Any element of the algebra of complex quaternions $\mathbb{C}\otimes\mathbb{H}$ can be written as a real linear combination of the following eight elements $1$, $i$, $\hat{e_1}$, $\hat{e_2}$, $\hat{e_3}$, $i\hat{e_1}$, $i\hat{e_2}$ and $i\hat{e_3}$. The complex unit $i \in \mathbb{C}$ commutes with every element of the algebra with $i^2=-1$ and $\hat{e_1}$, $\hat{e_2}$, $\hat{e_3}$ are the quaternionic units denoted by $\hat{e_k} \in \mathbb{H}$ such that they obey the commutation relations
\begin{equation}
 \hat{e_k}\hat{e_j}=-\delta_{kj}+\epsilon_{kjl}\hat{e_l}.
\end{equation}
$i,j,k, . . . = 1,2,3$ and the symbol $\epsilon_{ijk}$ is totally antisymmetric with $\epsilon_{123} = +1$.
One can easily identify $i\hat{e_1}$, $i\hat{e_2}$ and $i\hat{e_3}$, respectively, with the usual Pauli matrices $\sigma_1$, $\sigma_2$ and $\sigma_3$.
 Let us consider\footnote{$*$ to the right of a symbol takes its complex conjugate. The bar over a symbol means its quaternion conjugate. Then $\dagger$ refers to Hermitian conjugation consisting of complex conjugation followed by quaternion conjugation or vice versa.} the (anti-Hermitian) co-frame 1-form
\begin{equation}
  e=ie^0+\sum_{k=1}^{3}e^k\hat{e_k}=-e^{\dagger}
\end{equation}
in terms of which we can express the metric of spacetime as
\ba
g=Re(e\otimes\bar{e}).
\ea
A local Lorentz transformation is induced by the action of a unit quaternion $Q \in SL(2,\mathbb{C})$ ($Q\bar{Q}=\bar{Q}Q=1$) that converts the co-frame according to
\begin{equation}
  e \rightarrow \hspace{2mm} QeQ^{\dagger}.
\end{equation}
The following are the definitions for the connection, torsion, and curvature forms over space-time:
\ba
\omega &=&\sum_{k=1}^{3}\omega^k\hat{e_k}, \nonumber \\
T&=&de + \omega\w e + e \w \omega^{\dagger}=iT^0+\sum_{k=1}^{3}T^k\hat{e_k}=-T^{\dagger}, \nonumber \\
R&=&d\omega+\omega \w \omega = \sum_{k=1}^{3}R^k\hat{e_k},
\ea
with
\ba
\omega^k=-\frac{1}{2}(i\omega^0_{\;\;k}+\frac{1}{2}\epsilon_{ijk}\omega^{i}_{\;\;j}), \quad
R^k=-\frac{1}{2}(iR^0_{\;\;k}+\frac{1}{2}\epsilon_{ijk}R^{i}_{\;\;j}).
\ea
 It should be noted from above that $T$ is anti-Hermitian so the components $(T^0,T^1,T^2,T^3)$ are real 2-forms.
Moreover $\omega$ and $R$ are $SL(2,\mathbb{C})$ valued; consequently $\omega^k$ and $R^k$ are complex.
The Bianchi identities follow from the structure equations above as their integrability conditions:
\ba
dT + \omega \wedge T -T \wedge \omega^{\dagger} = R \wedge e - e \wedge R^{\dagger},
\ea
and
\ba
dR + \omega \wedge R - R \wedge \omega = 0.
\ea
It is not difficult to verify the following local Lorentz transformation rules
\ba
\omega \rightarrow Q \omega \bar{Q} + Qd\bar{Q}, \quad  T \rightarrow O T Q^{\dagger}, \quad R \rightarrow Q R \bar{Q}.
\ea

\noindent A Weyl spinor transforms as $\phi \rightarrow Q\phi$. Then we define its  covariant exterior derivative
\ba
\nabla \phi = d\phi + \omega \phi
\ea
that transforms properly under  local Lorentz transformations as $\nabla \phi \rightarrow  Q \nabla \phi$.  Similarly for a conjugate (dotted) Weyl spinor
$\dot{\phi} \rightarrow \dot{\phi} Q^{\dagger}$ we have
\ba
\nabla\dot{\phi} = d \dot{\phi} - \dot{\phi} \omega^{\dagger}.
\ea
These Weyl spinors can be represented  in the algebra of complex quaternions  by left ideals generated by $L_1:(U^1,U^2)$, $L_2:(W^1,W^2)$
or the right ideals generated by $R_1:(U^1,W^2)$ and $R_2:(U^2,W^1)$
where
\ba
U^1=\frac{1}{\sqrt{2}}(1+i\hat{e_3}), \quad U^2=\frac{1}{\sqrt{2}}(\hat{e_2}+i\hat{e_1}), \nonumber \\
W^1=\frac{1}{\sqrt{2}}(1-i\hat{e_3}), \quad W^2=\frac{1}{\sqrt{2}}(\hat{e_2}-i\hat{e_1}).
\ea

\noindent  An $SL(2,\mathbb{C})$ transformation applied to spinors can be induced by the left and right actions of a unit quaternion $Q$  according to
\begin{alignat}{4}
\phi&=\phi_1U^1+\phi_2U^2 &&\rightarrow&& \hspace{2mm} Q\phi \nonumber \\
\dot{\phi}&=\phi_{\dot{1}}U^2+\phi_{\dot{2}}W^1 &&\rightarrow&& \hspace{2mm} \dot{\phi} Q^{\dagger}.
\end{alignat}
By definition, a gravitino 1-form $\psi$  is an (odd-Grassmann) Weyl  spinor valued 1-form  that obeys the Majorana condition
\ba
\psi^{\dagger} = -i \dot{\psi}.
\ea

\noindent Finally, in terms of complex quaternion valued exterior differential forms, simple supergravity field equations  consist
of the coupled Einstein equation
\ba
R \wedge e = i \left ( \nabla \psi \wedge \dot{\psi} \right ),
\ea
Rarita-Schwinger equation
\ba
\bar{e} \wedge \nabla \psi = 0,
\ea
and the algebraic torsion
\ba
T = i \left ( \psi \wedge \dot{\psi} \right ) .
\ea

\bigskip

\section{Exact Colliding Wave Solutions}

\noindent We start with the Bell-Szekeres metric
\ba
g = 2 e^{-M} du dv + e^{-U}  \cosh W \left (e^{V} dx^2 + e^{-V}  dy^2 \right ) - 2 \sinh W  dx dy
 \ea
written in terms of complex null coordinates $u= \frac{z+t}{\sqrt{2}},  v=\frac{z-t}{\sqrt{2}},  \zeta =\frac{x+iy}{\sqrt{2}}$ where  $M,U,V,W$ are functions of $u,v$ only.
We work under the simplifying assumption\footnote{ We could have kept $V\neq0$ and $W\neq0$ at this point. But the Rarita-Schwinger equation forces the choice $V=W=0$.}
 $V = 0 = W$. Then we introduce the complex null co-basis 1-forms
$$
l = \frac{(e^3 +e^0)}{\sqrt{2}} = e^{-M/2} du  , \; n =\frac{(e^3 -e^0)}{\sqrt{2}}= e^{-M/2} dv, \;  m =\frac{(e^1 +ie^2)}{\sqrt{2}}=e^{-U/2} d\zeta;
$$
and consider the following ansatz for the gravitino 1-form
\ba
\psi = \xi_1(u,v) m U^1 + \xi_2(u,v) m^{*} U^2
\ea
 where $\xi_1(u,v)$ and $\xi_2(u,v)$ are complex (odd-Grassmannn valued) functions to be determined.
 The corresponding ansatz in the language of Majorana 4-spinor valued gravitino 1-form is established below in the Appendix.
 The space-time torsion 2-form turns out to be
 \ba
 T = \sqrt{2} m \wedge m^{*} \left ( -|\xi_1|^2 U^1 + |\xi_2|^2 W^1\right ).
 \ea
We substitute these in the Rarita-Schwinger field equation that reduces to the pair of equations
\ba
2 \frac{\partial \xi_1}{\partial v}  - \xi_1 \left ( \frac{\partial U}{\partial v}  + \frac{1}{2} \frac{\partial M}{\partial v} +\frac{i}{\sqrt{2}} e^{-M/2} |\xi_2|^2 \right ) = 0, \\
2 \frac{\partial \xi_2}{\partial v} - \xi_2 \left (  \frac{\partial U}{\partial u}  + \frac{1}{2} \frac{\partial M}{\partial u} -\frac{i}{\sqrt{2}} e^{-M/2} |\xi_1|^2   \right ) = 0. \nonumber
\ea
A formal integration yields the solution
\ba
\xi_1(u,v) = \varphi_1(u)  exp\left ( \frac{1}{2} (U+M/2)  +  \frac{i}{2 \sqrt{2}} \int^v e^{U(u,v^{\prime})} |\varphi_{2}(v^{\prime})|^2 dv^{\prime} \right )  ,
\ea   \ba
\xi_2(u,v) = \varphi_2(v) exp\left ( \frac{1}{2} (U+M/2)  -  \frac{i}{2 \sqrt{2}} \int^u e^{U(u^{\prime},v)} |\varphi_{1}(u^{\prime})|^2 du^{\prime} \right )  .
\ea
Let $\varphi_1(u)=|\eta_1| e^{i\alpha(u)}$ and $\varphi_2(v)=|\eta_2| e^{i\beta(v)}$ where $\eta_1,\eta_2$ are complex, odd-Grassmann valued constants.
$\alpha(u)$ and $\beta(v)$ are arbitrary phase functions to be fixed.
We now substitute in the Einstein field equation that reads
\begin{eqnarray}
\frac{\partial^2 M}{\partial u \partial v} + \frac{1}{2} \frac{\partial U}{\partial u} \frac{\partial U}{\partial v} + e^{2U} |\eta_1|^2 |\eta_2|^2 &=& 0, \\
\frac{\partial^2 U}{\partial u \partial v} - \frac{\partial U}{\partial u} \frac{\partial U}{\partial v} &=& 0, \nonumber \\
 \frac{e^{-U} }{2} \left (   \frac{\partial^2 U}{\partial u^2} +\frac{\partial M}{\partial u} \frac{\partial U}{\partial u} -\frac{1}{2} \left ( \frac{\partial U}{\partial u} \right )^2 \right )
&+&\sqrt{2} \frac{\partial \alpha}{\partial u} |\eta_1|^2  \nonumber  \\
 &+& \frac{1}{2} |\eta_1 |^2 |\eta_{2}|^2 \int^{v}  \frac{\partial U}{\partial u}  e^{U(u,v^{\prime})} dv^{\prime} =0,  \nonumber \\
   \frac{e^{-U} }{2} \left (   \frac{\partial^2 U}{\partial v^2} +\frac{\partial M}{\partial v} \frac{\partial U}{\partial v} -\frac{1}{2} \left ( \frac{\partial U}{\partial v} \right )^2 \right )
&-&\sqrt{2} \frac{\partial \beta}{\partial v} |\eta_2|^2 \nonumber \\
 &+& \frac{1}{2} |\eta_1 |^2 |\eta_{2}|^2 \int^{u}  \frac{\partial U}{\partial v}  e^{U(u^{\prime},v)} du^{\prime} =0.   \nonumber
  \end{eqnarray}

\noindent In order to solve these, we consider in particular two plane-fronted progressive super-waves that
 propagate along the $z$-axis, one to the left in Region:II where $ u>0, \; v<0$ and the other to the right in Region:III where $u<0,\; v>0$.
 Their wave fronts are both parallel to the $xy$-plane and look face to face as they propagate. They collide at the coordinate origin $(u=0,v=0)$ and continue  on their way
 along the $z$-axes in Region:IV where $u>0,\; v>0$.

\begin{figure}[H]
  \centerline{\includegraphics[scale=1]{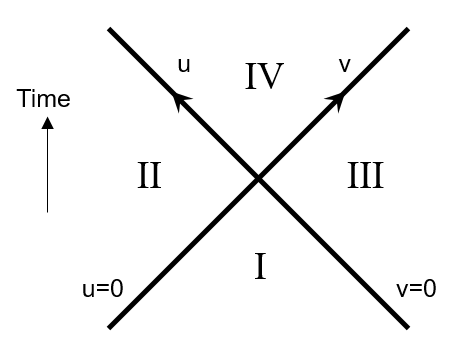}}
  \caption{Spacetime is divided into four regions as shown in the figure. Two spacelike coordinates have been suppressed.}
\end{figure}

\newpage
 We give below exact solutions in each of these regions  with the constants of integration arranged in such a way
 that solutions in separate regions connect consistently to those in other regions across the boundary null cone $u=0$ and $v=0$
 according to the O'Brien-Synge junction conditions:
 \begin{description}

 \item{Region:I, \; u<0, v<0. } Flat region.
 \ba
  M_{I}=U_{I} =0 , \; \; \psi_{I}=0.
  \ea
  \item{Region:II, \; u>0, v<0. } Left-moving super-wave.
  \ba
  M_{II}(u) =2 \alpha(u) + \ln (1 + \sqrt{2} |\eta_1|^2 u ) ,\; U_{II}(u) = -\ln (1 + \sqrt{2} |\eta_1|^2 u ),
\nonumber \\
 \psi_{II} = |\eta_1|e^{i\alpha(u)} e^{\frac{U}{2} + \frac{M}{4} - i\frac{U}{4}} U^1.
 \ea
   \item{Region:III, \; u<0, v>0. } Right-moving super-wave.
   \ba
   M_{III}(v) =-2 \beta(v) + \ln (1 + \sqrt{2} |\eta_2|^2 v ) ,\; U_{III}(v) = -\ln (1 + \sqrt{2} |\eta_2|^2 v ), \nonumber \\
 \psi_{III} = |\eta_2|e^{i\beta(v)} e^{\frac{U}{2} + \frac{M}{4} + i\frac{U}{4}} U^2.
 \ea
  \item{Region:IV, \; u>0, v>0. }  Interaction region.
\ba
M_{IV}(u,v) =2 \alpha(u) -2 \beta(v) + \ln (1 + \sqrt{2} |\eta_1|^2 u +\sqrt{2} |\eta_2|^2 v ) ,\nonumber \\ U_{IV}(u,v) = -\ln (1 + \sqrt{2} |\eta_1|^2 u +\sqrt{2} |\eta_2|^2 v ),
  \nonumber \\
 \psi_{IV} = \left (  |\eta_1|e^{i\alpha(u) - i\frac{U}{4}} U^1+ |\eta_2|e^{i\beta(v)+ i\frac{U}{4}}U^2 \right ) e^{\frac{U}{2} + \frac{M}{4} }.
 \ea
 \end{description}

\noindent The following non-zero Newman-Penrose connection coefficients are found:
\begin{eqnarray}
\rho &=&  -\frac{e^{ M/2 +U}}{\sqrt{2}} |\eta_2|^2 \theta(u)   \\
\mu &=&    \frac{e^{ M/2 +U}}{\sqrt{2}} |\eta_1|^2 \theta(v)       \nonumber \\
\gamma &=&     -\frac{e^M}{2} \left (   \frac{\partial \beta}{\partial v} \theta(v) + \beta \delta(v) + \frac{e^{U}}{\sqrt{2}} |\eta_1|^2 \theta(v)   \right )  \nonumber \\
\epsilon &=&       -\frac{e^M}{2} \left (   \frac{\partial \alpha}{\partial u} \theta(u) + \alpha \delta(u) + \frac{e^{U}}{\sqrt{2}} |\eta_2|^2 \theta(u)   \right )     \nonumber
\end{eqnarray}
Similarly the curvature scalars turn out to be
\ba
\Psi_0 = \Psi_1 =0, \quad \Psi_2 = -\frac{2}{3}e^{M+2U} \theta(u) \theta(v) |\eta_1|^2 |\eta_2|^2 .
\ea
Therefore our solution describes  geodesic, shear-free, non-rotating plane waves with  positive expansion in the interaction region.
Consequently a curvature singularity doesn't develop.

\noindent We calculate the following non-vanishing orthonormal components of the gravitino energy-momentum tensor (in Region IV):
\ba
T_{00} = T_{33} = e^{2\alpha -2\beta} \left ( |\eta_2|^2 \frac{\partial \beta}{\partial v} - |\eta_1|^2 \frac{\partial \alpha}{\partial u}  \right ), \nonumber \\
 \quad T_{03} = T_{30} = e^{2\alpha -2\beta} \left (  |\eta_2|^2 \frac{\partial \beta}{\partial v} +|\eta_1|^2 \frac{\partial \alpha}{\partial u}  \right ).
\ea
In particular for plane waves with $\alpha(u) =-ku\theta(u), \quad \beta(v)=kv\theta(v)$ where $k>0$ denotes the frequency, we have
\ba
 T_{00} = T_{33} =k e^{-2k(u+v)} (|\eta_1|^2 \theta(u) +|\eta_2|^2 \theta(v)),  \nonumber \\ T_{03} = T_{30} = k e^{-2k(u+v)} (-|\eta_1|^2 \theta(u)+|\eta_2|^2\theta(v)).
 \ea

\section{Concluding Remarks}

\noindent We gave a new  family of exact solutions to the simple supergravity field equations that describe the collision of two plane-fronted progressive gravitino waves
 that collide face-to-face in a non-Riemannian space-time setting with algebraic torsion.

\medskip

Both the metric tensor and the gravitino field are consistently patched across the null-cone by implementing the O'Brien-Synge boundary conditions\cite{obrien-synge}.
It turns out that an essential curvature singularity doesn't develop during such collisions.

\medskip

We also note that our family of solutions is non-trivial in the sense  they cannot be super-gauge generated from a vacuum Einstein configuration\cite{aichelburg-urbantke}.
Furthermore the gravitino energy-momentum tensor in the interaction region does not vanish, in contrast with the previously found family of  exact solutions of \cite{rosenbaum}.

\medskip

We used the language of complex quaternion valued exterior differential forms over the 4-dimensional space-time in our construction. 
First of all, the exterior differential forms are coordinate independent and all relevant tensors in this language remain free of indices $\mu,\nu,\dots$ in a coordinate chart $\{x^{\mu}\}.$
On the other hand, the algebra of complex quaternions carry representations of the spin cover of the local Lorentz group that can be used
to express  the local frames, connection and curvature forms without Lorentz indices $a,b,\dots$.  Thus an algebraic formulation of 4-dimensional space-time geometries is achieved with no indices. The massless chiral spinor fields are introduced in this language by considering minimal left ideals as spin basis.These correspond to undotted spinors in  the NP formalism while the dotted spinors of NP formalism are identified as minimal right ideals with a corresponding spin basis.The main advantage of our algebraic approach
is that once a local Lorentz frame is chosen, it automatically fixes the corresponding spin basis, thus the need to find
 a convenient set of $\gamma$-matrices  associated with the chosen local co-frame does not arise.

\medskip

The exact solutions given above might be useful in the context of some recent numerical results that point toward the formation of
black holes in the head-on collision of null particles\cite{pretorius-east}. We also wish to speculate that
the collision of massless gravitinos in supergravity may provide a fresh look at gravitational memory effects\cite{bieri}.

\bigskip

\section{Acknowledgement}
One of us (T.D.) thanks the Turkish Academy of Sciences (TUBA) for partial support.

\bigskip
\newpage

\section{Appendix}

\noindent We use the following Majorana realization of the $\gamma$-matrices
explicitly given by the real matrices
\begin{eqnarray}
\gamma_0= \left ( \begin{array}{cc} i\sigma_2 & 0 \\ 0 & -i\sigma_2  \end{array}  \right )&,&  \gamma_1=  \left ( \begin{array}{cc} \sigma_3 & 0 \\ 0 & -\sigma_3  \end{array}  \right ),  \gamma_2 =   \left ( \begin{array}{cc} 0& i\sigma_2  \\ -i\sigma_2 &0 \end{array}  \right ),  \nonumber \\
 \gamma_3=  \left ( \begin{array}{cc} \sigma_1 & 0 \\ 0 & \sigma_1  \end{array}  \right )&,&
\gamma_5= \gamma_0 \gamma_1 \gamma_2 \gamma_3 = \left ( \begin{array}{cc} 0& i\sigma_2 \\ i\sigma_2 &0 \end{array}  \right ).
\end{eqnarray}
A Majorana 4-spinor that is a self-charge conjugate 4-spinor in this realization turns out to be a real 4-spinor:
\ba
\lambda_{C} \equiv \mathcal{C} \bar{\lambda}^{T} = \lambda
\ea
where $\bar{\lambda} = \lambda^{\dagger} \mathcal{C}$ with $\mathcal{C} =\gamma_0$.

\medskip

\noindent  We make use of the one-to-one correspondence there exists between an (undotted) 2-component Weyl spinor
\ba
\varphi = \varphi_1 U^1 + \varphi_2 U^2
\ea
and a 4-component (real) Majorana 4-spinor
\ba
\lambda = \left ( \begin{array}{c}   {\varphi_{2}}^{\prime \prime} \\   {\varphi_{1}}^{\prime} \\ {\varphi_{1}}^{\prime \prime}  \\ -{\varphi_{2}}^{\prime} \end{array} \right )
\ea
where  $\varphi_{J} =  \varphi_{J}^{\prime} + i \varphi_{J}^{\prime \prime}, J=1,2.$
Then
given the gravitino 1-form  in the language of complex quaternions as
\ba
\psi = \xi_1(u,v) m U^1 + \xi_2(u,v) m^{*} U^2 ,
\ea
we substitute in the expressions $m = \frac{1}{\sqrt{2}} ( e^1 + ie^2 ), \quad  m^{*} = \frac{1}{\sqrt{2}} ( e^1 - ie^2 )$
and
 after some algebra obtain the expansion
\ba
\psi = \frac{1}{\sqrt{2}} \left [(  {\xi_{1}}^{\prime} +i \xi_{1}^{\prime \prime}  ) U^1 + ( \xi_{2}^{\prime} +i \xi_{2}^{\prime \prime}  ) U^2 \right ] e^1
+  \frac{1}{\sqrt{2}} \left [( -\xi_{1}^{\prime \prime} + i\xi_{1}^{\prime}    ) U^1 + ( \xi_{2}^{\prime \prime} -i \xi_{2}^{\prime}  ) U^2 \right ] e^2.  \nonumber
\ea
Therefore in the language of Majorana 4-spinors, our gravitino 1-form is given by
\ba
\psi = \psi_a e^a = \psi_1 e^1 + \psi_2 e^2
\ea
where
\ba
\psi_1 = \frac{1}{\sqrt{2}} \left ( \begin{array}{c} \xi_{2}^{\prime \prime} \\   \xi_{1}^{\prime} \\ \xi_{1}^{\prime \prime}  \\ -\xi_{2}^{\prime}   \end{array} \right ), \quad
\psi_2 = \frac{1}{\sqrt{2}} \left ( \begin{array}{c} -\xi_{2}^{\prime} \\   -\xi_{1}^{\prime \prime} \\ \xi_{1}^{\prime}  \\ -\xi_{2}^{\prime \prime} \end{array} \right ). \nonumber
\ea

\end{document}